\documentclass[twocolumn,pre]{revtex4-1}
\usepackage{graphicx,xcolor}
\usepackage{pgfplots}
\usepackage{epstopdf}
\usepackage{amsmath,amsfonts,amssymb,paralist,mathrsfs}
\usepackage{bm}
\usepackage{color}
\usepackage{verbatim}
\usepackage{enumerate}
\begin{document}
\title{Non-anomalous diffusion is not always Gaussian}

\author{Giuseppe~Forte$^{1}$, Fabio~Cecconi$^{2}$,
Angelo~Vulpiani$^{1,3}$}
\address{$^1$ Dipartimento di Fisica Universit\`a di Roma ``Sapienza'',
Piazzale Aldo Moro 2, I--00185 Roma, Italy\\
$^2$ CNR--Istituto dei Sistemi Complessi (ISC), Via dei Taurini 19,
I--00185 Roma, Italy\\
$^3$ CNR--Istituto dei Sistemi Complessi (ISC), Piazzale Aldo Moro 2,
I--00185 Roma, Italy.}

\begin{abstract}
Through the analysis of unbiased random walks on fractal trees and
continuous time random walks, we show that even
if a process is characterized by a mean square
displacement (MSD) growing linearly with time (standard behaviour)
its diffusion properties can be not trivial.
In particular, we show that the following scenarios are consistent
with a linear increase of MSD with time:
i) the high-order moments, $\langle [x(t)]^q \rangle$ for $q>2$
  and the probability density of the process exhibit multiscaling;
ii) the random walk on certain fractal graphs, with non integer spectral
    dimension, can display a fully standard diffusion;
iii) positive order moments satisfying standard scaling do not imply
     an exact scaling property of the probability density.
\end{abstract}

\maketitle

\section{Introduction \label{sec:intro}}
A colloidal particle in a fluid at thermal equilibrium 
undergoes a random displacement $x(t)$ due to 
the collisions with the surrounding molecules. Einstein~\cite{Einstein} 
proved for the first time that the random variable $x(t)$ 
follows a Gaussian distribution at large times  (see~\cite{FreyandKroy} 
and reference therein for a modern perspective). 
In particular, the mean square displacement (MSD) is proportional to 
the elapsed time, i.e.
$$
\langle x^{2}(t)\rangle\sim t
$$
$\langle\ldots\rangle$ being the ensemble average
over a set of initial conditions $x_{i}(0)$ $(i=1,2,3,\ldots)$. 
This scenario is referred to as {\em normal diffusion} and it 
is a directed consequence of the Central Limit Theorem (CLT), as the 
total displacement of a given particle at time $t$ 
$$
x_i(t)=\sum_{s=0}^{t}\Delta x_i(s)
$$
is the sum over the elementary displacements, 
$\Delta x_{i}(s) = x_i(s) - x_i(s-1)$. 
A classical result of the probability theory states 
that if $\Delta x_i(s)$ are independent (or weakly dependent) variables 
then $x_{i}(t)$ is a Gaussian-distributed variable at large $t$.

It is easy to show that a linear behaviour of the MSD is not a prerogative 
of Gaussian distribution only, in fact every distribution with a
self similar asymptotic property 
\begin{equation}
P(x,t) \sim \frac{1}{\lambda(t)} f\bigg(\frac{x}{\lambda(t)}\bigg)
\label{eq:simplescale}
\end{equation}
satisfies the condition $\langle x^{2}(t) \rangle \sim t$, 
upon choosing the lengthscale such that $\lambda(t) \sim t^{1/2}$. 
Therefore, the knowledge of $\langle x^{2}(t) \rangle$ alone
is poorly informative and its linear behaviour 
is not sufficient to assess neither that  
the diffusion is standard nor that the corresponding 
distribution $P(x,t)$ is Gaussian, see Ref.~\cite{Granick012} for a 
nice and interesting analysis. 

Moreover, deviations from the standard behavior are now well known and 
frequently observed in experiments, computer simulations, 
natural, economic and social processes \cite{klages_AnTrBook}. 
Such deviations are classified as anomalous diffusion 
\begin{equation}
\langle x^{2}(t)\rangle \sim t^{2\nu}
\label{eq:msd_anomal}
\end{equation}
with $\nu\ne 1/2$ \cite{Review90,PhysRep2000,RepProgPhys2013}. 
The case $\nu<1/2$ is called subdiffusion 
whereas $\nu>1/2$ is known as enhanced diffusion or superdiffusion.
For a process characterized by anomalous diffusion, the   
simplest scenario that can occur is a corresponding distribution 
which, for enough large $t$, 
still satisfies Eq.~(\ref{eq:simplescale}) with 
$\lambda(t) \sim t^{\nu/2}$. 

The self similar scaling~(\ref{eq:simplescale}) of the PdF 
automatically establishes the specific relationship
\begin{equation}
\langle |x(t)|^q \rangle = 
\langle x^{2}(t)\rangle^{q/2} \sim \lambda(t)^{q/2} \;.
\label{eq:magicmoments}
\end{equation}
among the moments in both anomalous and standard regimes. 

More interesting situations occur when the property 
\eqref{eq:magicmoments} is violated and it is replaced by the 
more general behaviour 
\begin{equation}
\langle |x(t)|^q \rangle \sim t^{q\nu(q)}
\label{eq:stranomal}
\end{equation}
where $q\nu(q)$ is a nonlinear function of $q$.
This is the case of a superdiffusive process that could be 
affected by the so-called {\em strong anomalous diffusion} 
\cite{Castiglione,Andersen}. 
The property \eqref{eq:stranomal} 
is generally referred in the literature to as {\em multiscaling} 
\cite{Multiscale1,Multiscale2,Multiscale3} in order to 
distinguish it from the {\em ordinary} or {\em simple scaling} 
characterized by self similarity, Eq.~(\ref{eq:simplescale}).

Indeed, the presence of a non-constant spectrum of exponents 
$\nu(q)$ for the time behaviour 
of the moments implies that the self similarity 
(\ref{eq:simplescale}) fails, so there is no chance to have a unique 
collapse of the PdFs at different times onto a single curve. 
The simplest, yet not unique possibility, occurs 
when the bulk and the tails of a PdF satisfy a different scaling law 
thus undergoing two separate collapses.
 
In this paper, we are interested in discussing through examples 
how the scaling behaviour~(\ref{eq:simplescale}) can be satisfied or violated. 
This issue becomes rather crucial in single-particle tracking 
experiments~\cite{RepProgPhys2013,Saxton1997}, when from a long but finite 
dataset of position measurements one would infer the statistical properties of
the underlying particle dynamics.  

Specifically, we will show that even in the apparently safe cases 
where $\nu(2)=1/2$, a non Gaussian PdF is possible with  
anomalous scaling of moments of order $q>2$, in particular 
$\langle x^4(t) \rangle \sim t^{4\nu(4)}$ with $4\nu(4)>2$.    
Conversely, there are cases with a standard scaling of moments
without a collapse of the PdF. 
Actually this is not surprising, as the knowledge 
of all the positive integer moments is not {\em fully equivalent} to the 
knowledge of the PdF, indeed examples can be constructed where 
two different distributions share the same moments.
Hence the reconstruction of the probability distribution from the   
sequence, $\mu_k = \langle x^k\rangle$ ($k=0,1,2,\ldots,\infty$),   
of positive integer moments is a classical and delicate 
issue of statistical mathematics, known as 
{\em the problem of moments}, which was formulated by T. Stieltjes in 1894 
\cite{Akhiezer65}. 
A general solution was given by Carleman \cite{Carleman1922} who 
identified a sufficient condition for which a probability distribution 
is uniquely determined by its infinite sequence of positive integer moments.

The paper is organized as follows: in Sec.~2, 
we revisit the continuous time random walk as an example of strong anomalous 
behaviour showing that, despite the condition $\nu(2) = 1/2$, its PdF has 
a multiscaling structure and $\nu(4)>1/2$.
In Sec.~3 and Sec.~4, we study the properties of the random walk 
on two fractal-like tree structures called in the following 
Nice Tree of dimension $k$ ($NT_k$) and Super Nice Tree ($SNT$) respectively.

The random walk on both graphs exhibits an ordinary scaling of moments,  
however, while in the $NT_k$ we find also a self similar scaling  
of the PdF, the scaling of the PdF in the $SNT$ structure is 
violated at small arguments.
Conclusions are found in Sec.5.

\section{The diffusive properties of the 
Continuous Time Random Walk \label{sec:CTRW}}
Continuous time random walk (CTRW) was introduced by Montroll and Weiss, 
in a series of pioneering papers on the diffusion processes on lattices, 
see e.g. \cite{CTRW_latt,Weiss}. 
We consider here a CTRW variant known as velocity 
model~\cite{classific_CTRW}, where a 
particle undergoes a series of kicks (collisions) at random times 
$t_1,t_1,\ldots,t_n,\ldots$ and 
between two consecutive collisions its velocity remains constant. 
The position of the particle at time $t$, such that $t_{n} < t \leq t_{n+1}$, 
will be 
\begin{equation}
x(t) = x(t_{n}) + v(t_n)(t-t_{n})
\label{eq:CTRW}
\end{equation}
the time intervals $\tau_n = t_{n+1} - t_{n}$ are independent random variables 
with a truncated power law distribution 
\begin{equation}
P(\tau) \propto 
\begin{cases}
\tau^{-g} & \quad  1 \leq \tau \leq T \\
0         & \quad  \mbox{elsewhere}
\end{cases}
\end{equation}
with $g > 1$ and $v_{n} = \pm 1$ with equal probability. 
The lower  cutoff $t_{c} = 1$ is a regolarization to avoid the 
singularity from infinitesimally short steps, moreover we also 
introduce an upper cutoff $T$ whose technical utility will be clear below. 
The presence of cutoff $T$ implies that the hypothesis of the CLT 
for the process~(\ref{eq:CTRW}) are 
fulfilled, thus as $t\gg T$, it converges to a Gaussian process. 
However if $T$ is chosen sufficiently large \cite{TrunkLevy}, this 
convergence can be made slow enough that a long and robust 
pre-asymptotic regime of strong anomalous diffusion can be clearly 
observed. 
A quantity that will be important in the following is the 
$q$-order moment of the waiting time $\tau$ whose asymptotic 
scaling for large $T$ is the following
\begin{equation}
\langle\tau^q\rangle_c \sim  
\begin{cases} 
T^{1-g+q}    & \mbox{if~} q > g-1 \\
a(q,g)       & \mbox{if~} q < g-1 
\end{cases} 
\label{eq:tau_q}
\end{equation}
where $a(q,g)$ is a constant independent of $T$ and   
the index $c$ indicates the average over the ``truncated'' 
distribution. 

At $T\to \infty$, various diffusive regimes occur depending 
on the value of the exponent $g$, see Andersen et al.~\cite{Andersen}; 
in particular the case $g \in (3,4]$ corresponds to the anomalous diffusion, 
$\langle |x(t)|^{q}\rangle\sim t^{q\nu(q)}$, with
\begin{equation}
\label{eq:g_scaling}
q\nu(q)=
\begin{cases} 
q/2  , & q=1,2 \\ 
q+2-g, & q=3,4,5,\cdots
\end{cases}
\end{equation}
The above behaviour of $q\nu(q)$ is quite peculiar as it coincides with 
that one of standard diffusion for $q<2$, while for larger $q$, 
we have $q\nu(q) \ne q/2$, and this represents an example of strong anomalous 
regime. 

Following the reasoning of Andersen and coworkers~\cite{Andersen}, we 
compute the $q$-order moments of the variable 
\begin{equation}
x(t) = \sum_{i=1}^{n_t} v_i \tau_i 
\label{eq:displa}
\end{equation}
at different  $q$, where $n_t$ is the stochastic process counting the number 
of ``collisions'' (flights) the particle underwent within the time $t$. 
Odd-order moments $\langle [x(t)]^q \rangle$ vanishes for the symmetry 
$v \to  -v$ of the velocity distribution.
Even-order moments are nonzero and can be evaluated exploiting 
the following properties:\\ 
$\langle v_i \tau_j\rangle = 0$, 
$\langle v_i v_j \rangle = \delta_{ij}$, 
$\langle \tau_i \tau_j\rangle =  \langle\tau^2\rangle \delta_{ij}$. 
The lowest non-zero moments of $x(t)$ 
$$
\langle x^{q}(t)\rangle = 
\bigg \langle \bigg(\sum_{i=1}^{N} v_i \tau_i\bigg)^q \bigg\rangle 
$$
can be explicitly derived from the general multinomial formula
\begin{equation}
\langle x^{q}(t)\rangle  
= \sum_{\{\mathbf k\}}
\frac{q!}{k_1!k_2!...k_{N}!} 
\prod_{j=1}^{N}\langle (v_j \tau_j)^{k_j} \rangle
\label{eq:momq}
\end{equation}
with $\{\mathbf k\} = \{k_1\ldots k_N\}$ indicating the sets  
of non-negative even integer arrays such that 
$k_1 + k_2 +\ldots + k_{N} = q$.

In the above expression, we have implicitly taken the average 
$\langle n_t \rangle = N$ and exploited independence of the 
variables.  
\begin{table*}[t!]
\begin{tabular}{lc}
q      &  Moment  \\
\hline\hline
2  &  $ N \langle\tau^2\rangle$ \\
4  &  $ N \langle\tau^4\rangle + 3 N(N-1) \langle\tau^2\rangle^2 $ \\
6  &  $N\langle\tau^6\rangle +            
        15 N(N-1) \langle\tau^2 \rangle\langle\tau^4\rangle +  
        15 N(N-1)(N-2)\langle \tau^2 \rangle^3  $\\
8  &  $ N \langle\tau^8\rangle + 28 N(N-1) \langle\tau^6\rangle 
       \langle\tau^2\rangle + 35 N(N-1) \langle\tau^4\rangle^2 +  
     210 N(N-1)(N-2) \langle\tau^2\rangle^2\langle\tau^4\rangle +   
     105 N(N-1)(N-2)(N-3) \langle\tau^2\rangle^4$  
\end{tabular}
\caption{\label{tab:mom} 
Lowest four non-vanishing moments of the displacement Eq.~(\ref{eq:displa})}
\end{table*} 
In Table~\ref{tab:mom}, the explicit expressions of the lowest four 
non-vanishing moments are reported. With simple considerations 
it is clear what are the terms of the expansion 
which maximally contribute to the moments. 
First we have to observe that, when the time $t$ is so large that 
enough collisions (steps) occurred, $n_t\gg1$, we have in a good
approximation  $t\simeq \sum_{i=1}^{n_t} \tau_i$, 
therefore as a consequence of the Law of Large Numbers, 
$t \simeq N \langle\tau\rangle_c$, where 
$N = \langle n_t \rangle $ is  mean value of the number of time 
steps necessary to reach the time $t$. Accordingly 
for large times we can replace in all the formulas
$N \sim t/\langle\tau\rangle_c$, with 
$g$ such that $\langle\tau\rangle_c$ is finite as $T\to\infty$. 
Now two regimes have to be considered depending on the ratio $t/T$. 
For $t/T\ll 1$ (i.e. $t$ in the anomalous regime), the $q$-order moment is 
dominated by $N \langle\tau^q\rangle_c$, the largest term in $T$, then   
$$
\langle x^{q}(t)\rangle  
 \sim t \frac{\langle\tau^q\rangle_c} 
{\langle\tau\rangle_c}.
$$ 
Whereas, when  $t/T\gg 1$ ($t$ in the Gaussian regime)  
$$
\langle x^{q}(t)\rangle \sim t^q
\bigg(\frac{\langle\tau^2\rangle_c}{\langle\tau\rangle_c}\bigg)^q.
$$
finally, at $t\sim T$ there is the crossover from the anomalous to the 
standard behaviour. 
The exponent $q\nu(q)$ is determined by matching, at the cutoff $t\sim T$, the 
anomalous scaling of the q-moments $T^{q \nu(q)}$ with the scaling of 
the most divergent term in expansion (\ref{eq:momq}), 
$T\langle\tau^q\rangle_c/\langle\tau\rangle_c$ 
$$
T^{q \nu(q)} \sim T \times T^{1-g+q}  
$$
from which $q\nu(q) = 2 - g + q$ [see Eq.~(\ref{eq:g_scaling})].

\begin{figure}[h!]
\includegraphics[clip=true,keepaspectratio,width=0.45\textwidth]
{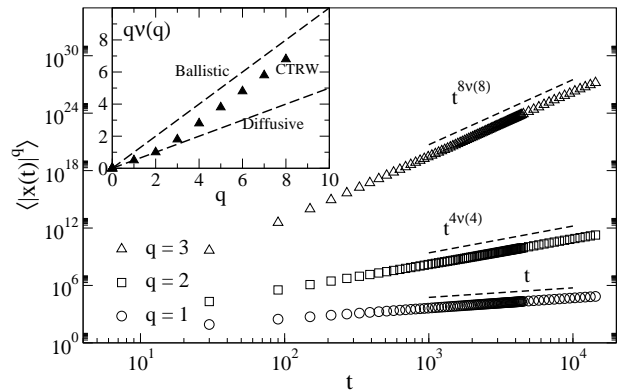}
\caption{\label{fig:Mom} $\langle x^{q}(t)\rangle$ vs $t$ $(q = 2,4,8)$ 
resulting by a simulation of $6.4\times 10^7$ CTRW realizations (\ref{eq:CTRW}) 
with $g = 3.2$; the inset shows the relation $q\nu(q)$ vs $q$ which
underlines the strong anomalous diffusive character of this particular system.}
\end{figure}

The behaviour $\langle |x(t)|^{q} \rangle \sim t^{q\nu(q)}$ of the 
CTRW moments is verified by simulating $6.4\times 10^7$ independent   
CTRW trajectories with $g=3.2$ and computing the ensemble 
average $\langle\cdots\rangle$ at different times.  
The results are reported in Fig.~\ref{fig:Mom}. 
The inset shows the nonlinearity of the 
exponent $q\nu(q)$ as a function of $q$ proving the multiscaling character 
of the diffusion.     

The relation (\ref{eq:g_scaling}) provides strong indications on 
the possible form of the PdF; the lowest order moments behave in time as in 
the case of normal diffusion, thus we expect that the $P(x,t)$ has a 
Gaussian bulk which scales as $P(x,t)=t^{-1/2} f(x/t^{1/2})$), with 
$f(u)=(2\pi\sigma^2)^{-1/2} \exp[-u^2/(2\sigma^2)]$ 
for moderate value of the 
argument $|u/\sigma|$. In the range $\sigma\sqrt{t}\ll x \lesssim c \sim t $ the 
behaviour of the high order moments suggests the following form of the PdF
\begin{equation}
\label{eq:CTRW_PdF}
P(x,t) =
\begin{cases} 
\displaystyle
\frac{1}{t^{1/2}} f\left(\frac{x}{t^{1/2}}\right), & x\le c \sim t 
\\ 
0,                                                 & x > c.
\end{cases}
\end{equation}  
The assumption (\ref{eq:CTRW_PdF}) is consistent with 
Eq.~(\ref{eq:g_scaling}) only if $x$ around $c \sim t$, the function 
$f$ is such that 
$$
f\left(\frac{x}{t^{1/2}}\right)\sim \left(\frac{x}{t^{1/2}}\right)^{-\alpha}
$$
namely, the tails decay as power-law behavior with an exponent $\alpha$ 
related to $g$. Let $x^* \sim \sigma\sqrt{t}$ denote the value at which the 
crossover between  
\begin{eqnarray*}
\displaystyle
\langle x^{q}(t)\rangle  
= \int_{0}^{\tilde{x}}dx x^{q}P_{t}(x) \sim   \\
\displaystyle
\int_{0}^{x^{*}}dx\;\frac{x^{q}}{t^{1/2}}f\left(\frac{x}{t^{1/2}}\right) 
+ \mbox{cost} 
\int_{x^{*}}^{c}dx\;\frac{x^q}{t^{1/2}}\left(\frac{x}{t^{1/2}}\right)^{-\alpha}
\end{eqnarray*}
the first term behaves as $t^{q/2}$, whereas the second one behaves as 
$t^{q +\frac{1}{2}-\frac{\alpha}{2}}$. Therefore for small $q$ the dominant 
contribution comes from the first term,  
$\langle x^{q}(t)\rangle \sim t^{q/2}$,  while for large $q$ the leading 
contribution is given by the second term. 
The scaling exponent $q +(1-\alpha)/2$ is in agreement 
with $q\nu(q) = q+2-g$, second line of
Eq.~(\ref{eq:g_scaling}), only if $\alpha = 2g - 3$ 
which is the exponent of the expected power-law behavior of tails 
outside the Gaussian bulk.  

\begin{figure}[h!]
\includegraphics[clip=true,keepaspectratio,width=0.45\textwidth]
{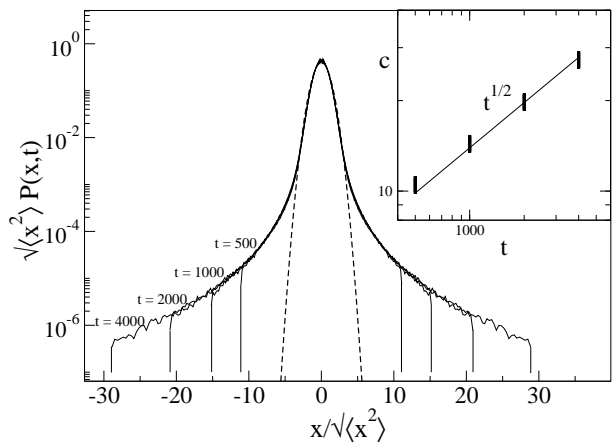}
\caption{\label{fig:Dist_CTRW} Rescaled probability densities of the CTRW
at different times for $g=3.2$. The PdF is obtained from the histogram 
over $6.4 \times 10^7$ CTRW positions at times $t=500, 1000, 2000, 4000$. 
The vertical lines are guide for the eyes to mark
the bounded support, $|x|\le c(t)$, of the distributions according to 
Eq.~(\ref{eq:CTRW_PdF}), the dashed line represents the Gaussian PdF.
The inset shows the scaling $c(t)/\sqrt{t}\sim t^{1/2}$
in Eq.~(\ref{eq:CTRW_PdF}).}
\end{figure}

The collapse of the rescaled PdF of the CTRW~(\ref{eq:CTRW}) 
at different times is shown in Fig.~\ref{fig:Dist_CTRW}.
As it discussed in the above argument, the PdF must have a bounded support, 
$|x|\lesssim c(t)$ see Eq.~(\ref{eq:CTRW_PdF}).   

It is worth remarking that the above result is not actually a violation 
of the Central Limit Theorem (CLT), as in the bulk all the PdFs collapse 
onto a Gaussian and only the far tails deviate from
this behavior. The CLT, indeed, does not grant anything on the nature of the 
tails, it only specifies the shape of the limit distribution within the 
scaling region, $|x(t)/\sqrt{t}| \sim O(1)$. The tails
outside such a  bound are not universal and generally not Gaussian. 
Analogously, there is no reason for the high order moments, 
which receive the main contribution from tails, to converge to 
the Gaussian moments.

We conclude this section with a methodological comment. 
The statistical properties of the CTRW like moments or PdF at a certain 
time $t$ have been computed by following an ensemble of walkers up to 
$t$. However we checked that results obtained from the ensemble 
of trajectories are perfectly consistent with those obtained using a long 
single-realization~\cite{Sokolov12}. 
The discrepancy between ensemble and 
single-trajectory analysis has important physical implications
as it may indicate deviations from standard 
Brownian diffusion because of ergodicity breaking and 
aging in the process dynamics \cite{Nonergodicity}. 
Here, we only mention this crucial issue 
while referring the reader to the works by Sokolov \cite{Sokolov12} 
and Barkai et al. \cite{Barkai_PhysToday} for a plain and nice
discussion. 
However we stress that the considered velocity CTRW, for $g>3$ 
has a finite average time $\langle t \rangle$ and the presence of a 
characteristic time scale in the dynamics excludes ambiguous results
from the single trajectory analysis and grants the equivalence
between ensemble and running averages.

\section{Diffusion on branched graphs}
A undirected graph is a collection of vertices pairwise connected, or not, 
by links. 
To each graphs of $N$-vertices, we can associate a $N\times N$ 
matrix A (adjacency matrix), such that, A$_{ij}$ = $1$ 
if there is an link between vertices $i$ and $j$, A$_{ij}$ = $0$ otherwise. 

An unbiased random walk on a graph can be defined in a natural way: 
a walker at time $t$ on the node $i$ can jump at time $t+1$ 
on the node $j$ only if A$_{ij}$ = $1$, with a transition probability 
$P\{i\to j\} = 1/n_i$, where $n_i = \sum_j\mbox{A}_{ij}$ is the number of links 
established by the node $i$. 
Then the diffusing variable is the position of the walker 
which takes on discrete values defined by the vertices of the graph. 
The latter assumption of equal-probability 
of the transition to nearest neighbors of $i$ can be relaxed.

The diffusion properties of a random walk on a graphs 
depend on both the {\em fractal dimension} \cite{Mandelbrot} 
$d_f$ 
and the {\em spectral dimension} $d_s$ 
\cite{AlexanderandOrbach}.  
The fractal dimension is related to the scaling of the number of points
in a sphere of radius $\ell$: $N(\ell) \sim \ell^{d_f}$; 
the spectral dimension is defined by the return probability $P_{t}(x)$ to a 
generic site $x$ in $t$ steps $P_{t}(x) \sim t^{-d_s/2}$.
The ratio between $d_s$ and $d_f$ determines the mean square displacement
through the relation \cite{Ben-Avraham} 
\begin{equation} 
\langle x^{2}(t)\rangle\sim t^{d_s/d_f}.
\label{eq:msd_graph}
\end{equation}

Analogously to the CTRW, 
we can investigate the behaviour of high-order moments and the 
possible collapse of the PdF.
In particular, we focus on the random walks on a class of graphs  
the {\em Nice Trees of dimension} $k$ ($NT_{k}$), that are recursive 
fractal trees with the remarkable property that 
\begin{equation}
d_{f} = d_{s}(k) = 1 + \frac{\ln{k}}{\ln{2}},
\label{eq:NTD_ds}
\end{equation}
i.e. the fractal and 
the spectral dimension coincide \cite{BurioniCassi,BurioniCassi2} for any 
value of $k$. 
Hence, despite their fractal-like structure, 
Eq.~\eqref{eq:msd_graph} prescribes a standard diffusion 
of the random walks on any $NT_k$. 
\begin{figure}[h!]
\includegraphics[clip=true,keepaspectratio,width=0.23\textwidth]
{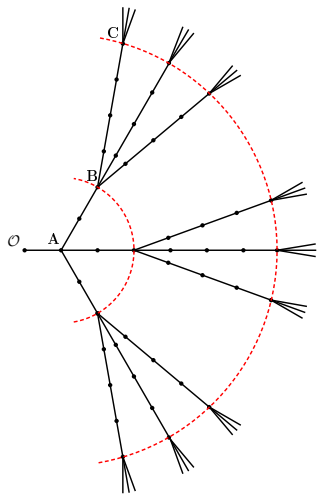}
\caption{Geometrical construction of a Nice Tree of dimension $k = 3$.
An origin $\mathcal{O}$ is connected with a point $A$ by a link of length 1;
from $A$ the tree splits in $k$ branches of length $2^{1}$ each.
The end point of each branch, in turn, splits again into $k$ branches of
length $2^{2}$. The procedure is recursively iterated.}  
\label{fig:NTD}
\end{figure}
A $NT_k$ graph is defined recursively as follows.  
An origin $\mathcal{O}$ is connected with a point $A$ by a link of length 1; 
from $A$ the tree splits in $k$ branches of length $2^{1}$ each. 
The end point of such branches, in turn, splits again into $k$ branches of 
length $2^{2}$ and so on (see Fig.~\ref{fig:NTD}). 

From the previous section, we learnt that a linear growth of the MSD in time 
(i.e. $2\nu(2)=1$) does not grant a Gaussian diffusive process 
thus, we can wonder about 
the consequences of the property \eqref{eq:NTD_ds} on 
the behaviour of the full spectrum of moments and the PdF.
In order to characterize the diffusion properties of the 
unbiased random walk on the $NT_{k}$, we need to 
numerically study the corresponding master equation of the process. 

\begin{figure}[h!]
\includegraphics[clip=true,keepaspectratio,width=0.45\textwidth]
{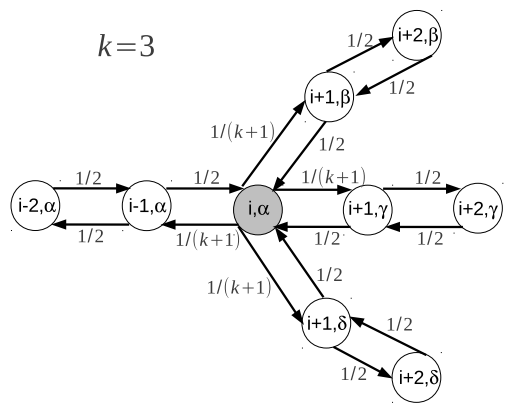}
\caption{\label{fig:map} Transitions around a branching point 
of a $NT_k$ with $k=3$ that have to 
be considered for the construction of the one-dimensional 
diffusion model.  
A walker on a branching point (shaded circle) 
has $k$ possibilities to make one step away from the origin and $1$ 
possibility to get one step closer.} 
\end{figure}

We can assign to each site the integer distance $x$, 
if it is connected to the origin 
$\mathcal{O}$ by the minimal path with $x$-links.  
Each sites of the $NT_k$ graph can be identified by a couple of 
indices $(x,\alpha)$ indicating the distance from $\mathcal{O}$  
(depth of the tree) and the corresponding branch (Fig.~\ref{fig:NTD}). 
The structure of the master equation governing 
the evolution of the probability $Q_t(x,\alpha)$ that a walker occupies 
at time t the state $(x,\alpha)$ is understood by considering the possible 
transitions around a generic branching point (Fig.~\ref{fig:map}). 
The equation  involves nearest neighbour sites $\{x-1,x,x+1\}$ which can 
belong to either the same branch or different but consecutive branches.
Formally, it can be written as  
\begin{eqnarray}
Q_{t+1}(x,\alpha) &=&W(x,\alpha|x-1,\alpha) Q_t(x-1,\alpha)  \nonumber\\
                  &+&\sum_{\beta} W(x,\alpha|x+1,\beta) Q_t(x+1,\beta)
\label{eq:mother_masteq}
\end{eqnarray}
with $\alpha,\beta$ identifying two consecutive 
branches and $W(x,\alpha|x-1,\alpha)$, $W(x,\alpha|x+1,\beta)$ are
the corresponding transition probabilities. 
As far as we are interested in the evolution of $P_{t}(x)$, 
the probability for the walker to be at distance $x$ at time $t$, 
we need to sum over the index $\alpha$, namely over all those
branches containing a site at distance $x$ from
the origin, then  $P_{t}(x) = \sum_{\alpha} Q_{t}(x,\alpha)$.
We can now 
consider the one-dimensional master equation 
obtained by ``projecting'' the complete master 
equation of the $NT_k$ onto the one-dimensional lattice 
$x \in \{n\}_{n=0}^{\infty}$.
The procedure is a generalization of that used in Ref.~\cite{RW_Bethe} 
and with reference to Fig.~\ref{fig:map}, it leads to the 
following three cases
\begin{eqnarray}
P_{t+1}(x-1) =& \displaystyle{\frac{1}{2} P_t(x-2) + \frac{1}{k+1} P_{t}(x)} 
\nonumber \\
P_{t+1}(x)   =& \displaystyle{\frac{1}{2} P_t(x-1) + \frac{1}{2} P_{t}(x+1)} 
\nonumber \\
P_{t+1}(x+1) =& \displaystyle{\frac{k}{k+1} P_t(x) + \frac{1}{2} P_{t}(x+2)} 
\label{eq:1D_NTD}
\end{eqnarray}
where the first and third equations holds only for branching points ($x=2^n-1$), 
the second one for all the other sites.   
Accordingly, the transition matrix $w(x\pm 1|x)$ of the one-dimensional RW  
from $x \to  x\pm 1$, in a time step reads  
\begin{equation}
 w(x+1|x) =
\begin{cases} 
\displaystyle{\frac{k}{k+1}},  &\mbox{if~}\; x=2^n-1\\ 
1/2,    &\mbox{elsewhere} 
\end{cases}  
\end{equation}
\begin{equation}
  w(x-1|x) = 
  \begin{cases}
 \displaystyle{\frac{1}{k+1}}, &\mbox{if~} x=2^n-1\\
 1/2,   &\mbox{elsewhere} 
\end{cases}
\end{equation}
where $2^n-1$ is the formula identifying the distance of the branching points 
from the origin $\mathcal{O}$,
$W(1|0) = 1$ is the condition for reflecting boundary in $\mathcal{O}$.   
The RW on NT$_{k}$ is thus mapped onto a RW on a one-dimensional lattice 
in a deterministic heterogeneous environment. 
The physical interpretation of the transition matrix is simple, 
if a walker sits on a branching point, there are $k$ possibilities to 
go one step away from origin and $1$ possibility to make one step closer.
Then the next step will take it either
farer from the origin with probability $p_{+}= k/(k+1)$ or closer the origin 
with probability $p_{-}= 1/(k+1)$. Whereas if the a walker is away from the 
branching point, both steps are unbiased, $p_{-} = p_{-} = 1/2$.   
The  
inhomogeneity stems from the branching points $x=2^{n}-1,\;n=1,2,3\cdots$ which 
represent special points (``defects'') but become exponentially rare as long as
the walker lies far away from the origin. Thus, far away from the origin,   
the process remains an unbiased RW for so long time that ``Gaussian 
character'' of the distribution is not altered by the presence defects.

It is well known that a coarse-graining procedure over a Markov process
generally spoils the Markov property. However as a consequence of the peculiar 
structure of the transition probabilities the reduction to the single  
``radial coordinate'' of Eq.~\eqref{eq:mother_masteq} produces
again a Markovian master equation.

The analytical solution $P_t(x)$ to master equation~\eqref{eq:1D_NTD} 
cannot be derived in a simple explicit form, however it  
can be easily obtained by iterating numerically Eq.~\eqref{eq:1D_NTD} 
from an initial condition. 
 
\begin{figure}[h!]
\includegraphics[clip=true,keepaspectratio,width=0.45\textwidth]
{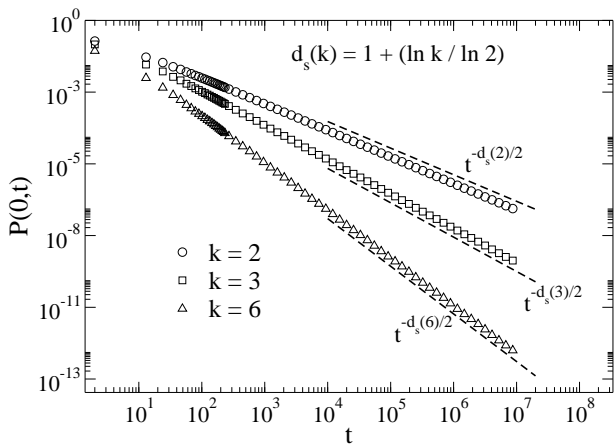}
\caption{\label{fig:Return_NTD}
Log-Log plot of the return probability to the origin in $t$ steps, 
$P_{t}(\mathcal{O})$ as a function of $t$,  for $NT_k$  
of dimension $k=2$ (circles), $k=3$ (squares) and $k=6$ (triangles).
Data are in agreement with the Graph-Theory prediction  
$P_{t}(\mathcal{O})\sim t^{-d_{s}/2}$ 
with $d_{s} = 1 + \ln(k)/\ln(2)$, Eq.~\eqref{eq:NTD_ds}, dashed lines.}  
\end{figure}

\begin{figure}[h!]
\includegraphics[clip=true,keepaspectratio,width=0.45\textwidth]
{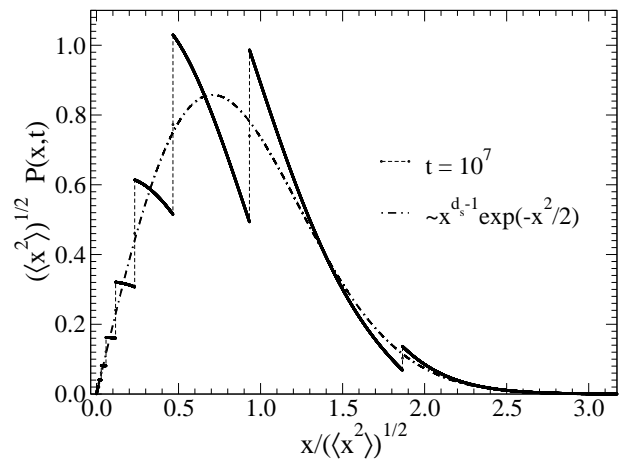}
\caption{\label{fig:Dist}
Probability density obtained by simulating the master equation
(\ref{eq:1D_NTD}) referring to a NT$_k$ with $k=2$. 
Dashed line indicates the analytical interpolation \eqref{eq:app_NTDpdf}.}
\end{figure}

As a first check of convergence of the numerical implementation of 
the 1D approach, we compute the return probability 
$P_{t}(\mathcal{O})$ to the origin in 
$t$ steps. 
Graph theory \cite{Ben-Avraham} predicts the power-law decay  
$$
P_{t}(\mathcal{O})\sim t^{-d_{s}(k)/2}\;
$$ 
with $d_{s}(k)$ from Eq.~\eqref{eq:NTD_ds}, data of Fig.~\ref{fig:Return_NTD}
perfectly verified the prediction.
Figure~\ref{fig:Dist} shows the simulation results for the 
probability density $P_{t}(x)$ 
rescaled to $x\to x/\sqrt{\langle x^{2}\rangle}$ and 
$P_{t}(x)\to\sqrt{\langle x^{2}\rangle}P_{t}(x)$ obtained by iterating 
the master Eq.~\eqref{eq:1D_NTD} for a NT$_k$ with $k = 2$.
The discontinuities of $P_{t}(x)$ are a clear signature of the ``defects'' 
(branching points on the tree) which interrupt the standard random walk in 
the passage from a branch to the successive one. 
The dash-dotted curve is the approximated solution  
\begin{equation}
F_t(x) = 
\frac{2 x^{d_s-1}}{\Gamma(d_s/2)(2t)^{d_s/2}} \exp(-x^{2}/2t)
\label{eq:app_NTDpdf}
\end{equation}
which well interpolates the exact numerical result. 
Expression~\eqref{eq:app_NTDpdf} is the radial Gaussian 
distribution in dimension $d_s(k)$ and it can be explained by 
the following heuristic argument.
Let $\tilde{P}_{t}\sim \exp(-x^{2}/2t)$ be the probability density at time 
$t$ of an unbiased one-dimensional RW from zero to infinity.
In a first approximation, $P_{t}(x)$ can be assumed as the product 
$N_{x}\tilde{P}_{t}(x)$, where $N_{x}$ is the number of sites 
at the same distance $x$ from the origin; we can write $N_{x}$ as $k^{n(x)}$, 
with $n(x)$ the number of branching points along 
a minimal-length path connecting $\mathcal{O}$ and $x$. 
The number $n(x)$ can be obtained by observing that the branching points are 
those located at $x_{br} = 2^{n} - 1$ $(n=1,2,3,\cdots)$, from which 
$n(x) = \lfloor \ln{(x+1)}/\ln{2}\rfloor$, so a given walker at distance 
$x$ has crossed $n(x)\approx \ln{(x+1)}/\ln{2}$ possible ramification points.
Now with the aid of Eq.~\eqref{eq:NTD_ds}, we can rewrite  
$N_{x} \sim x^{d_{s}(k)-1}$, which after normalization yields 
expression \eqref{eq:app_NTDpdf}. 

\begin{figure}[h!]
\includegraphics[clip=true,keepaspectratio,width=0.45\textwidth]
{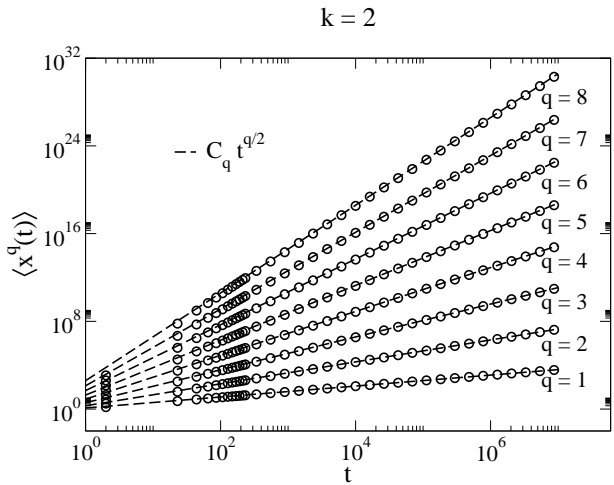}
\caption{
Time behaviour of the moments, $\langle x^{q}(t)\rangle$ ($q = 1,...,8$)
obtained by solving numerically the 1D master equation~\ref{eq:1D_NTD}, 
for a NT$_k$ with $k = 2$. All the moments show a Gaussian-like 
behaviour, which is well explained by the approximated PdF 
(\ref{eq:app_NTDpdf}). Dashed lines are the results~\ref{eq:approx_mom}.
}
\label{fig:approx_mom}
\end{figure}

Using the approximation $F_t(x)$, we can estimate all the 
moments $\langle x^{q}(t)\rangle$, 
\begin{equation}
\langle x^{q}(t)\rangle\approx
\int_{0}^{\infty}dx\;F_{t}(x) x^{q}=C_{q}t^{q/2}
\label{eq:approx_mom}
\end{equation}
where $C_{q}=2^{q/2}\Gamma(q/2 + d_{s}/2)/\Gamma(d_{s}/{2})$.
The agreement of formula~\eqref{eq:approx_mom} with the numerically
computed result is striking, see Fig.~\ref{fig:approx_mom}, 
considering that there is no free parameters. 

We can conclude this section noting that,    
despite the fractal complexity of NT$_k$, the RW on it does not 
develop a multiscaling character because its 
large-scale statistics remains Gaussian-like as clearly indicated 
by the shape of the approximated distribution (\ref{eq:app_NTDpdf}). 

In the next section, we modify 
the NT$_k$ geometry in order to achieve a RW process with 
Gaussian scaling of moments $\langle |x(t)|^q\rangle \sim t^{q/2}$ 
without Gaussian PdF.

\section{Diffusion on super - branched graphs}
\begin{figure}[h!]
\includegraphics[clip=true,keepaspectratio,width=0.25\textwidth]
{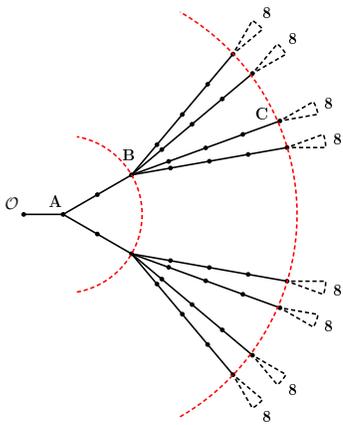}
\caption{\label{fig:SNTD} 
Sketch of the super branched Nice Tree with $k = 2$. It is 
recursively defined as a $NT_{k}$, but now each branch 
splits into $k^n$ new branches.}
\end{figure}

The structure of the NT$_{k}$ graph can be easily modified to 
generate a RW which exhibits standard scaling of all the moments 
$\langle |x(t)|^q\rangle\sim t^{q/2}$ without having a Gaussian PdF.
We change  $NT_{k}$ structure by defining a new type of graph, which we dub 
$SNT$ (Super Nice Tree).  
A $SNT$ (Fig.~\ref{fig:SNTD})  
is recursively defined as a $NT_{k}$, but at every branching site  
$x = 2^n - 1$ ($n = 1,2,3,...$) the tree splits in $k^n$ branches.

In analogy to the $NT_{k}$ case, if we are interested in the process of 
diffusion from the origin, we again have an equivalence 
with a random walk on the line with a deterministic 
distribution of defects (branching points). 
We have the transition matrix
\begin{equation}
 w(x+1|x) =
\begin{cases} 
 \displaystyle{\frac{k^{n}}{k^{n}+1}}, &\mbox{if~} x=2^n-1\\ 
 1/2,    &\mbox{elsewhere} 
\end{cases}  
\end{equation}
\begin{equation}
  w(x-1|x) = 
\begin{cases}
 \displaystyle{\frac{1}{k^{n}+1}}, &\mbox{if~} x=2^n-1\\
 1/2,   &\mbox{elsewhere} 
\end{cases}
\end{equation}
and the corresponding master equation
\begin{eqnarray}
P_{t+1}(x-1) =& \displaystyle{\frac{1}{2} P_t(x-2) + \frac{1}{k^n+1} P_{t}(x)} 
\nonumber \\
P_{t+1}(x)   =& \displaystyle{\frac{1}{2} P_t(x-1) + \frac{1}{2} P_{t}(x+1)} 
\nonumber \\
P_{t+1}(x+1) =& \displaystyle{\frac{k^n}{k^n+1} P_t(x) + \frac{1}{2} P_{t}(x+2)} 
\label{eq:1D_SNTD}
\end{eqnarray}

\begin{figure}[h!]
\includegraphics[clip=true,keepaspectratio,width=0.45\textwidth]
{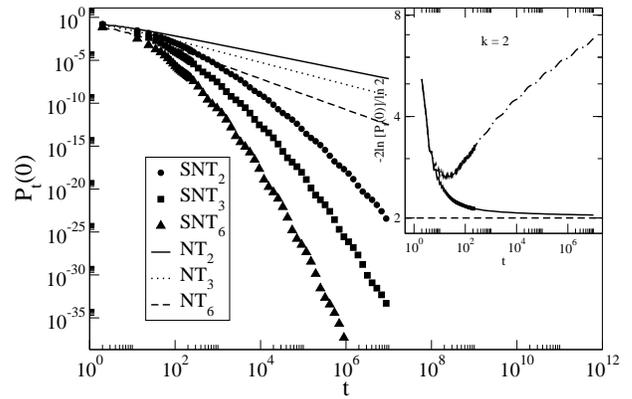}
\caption{\label{fig:return_SNTD}
Probability to come back to the origin in $t$ steps as function of time. 
The data shows the case of an $SNT$ ($k=2,3,6$) compared to the 
respective cases which one observe for $NT_{k}$. 
Inset: numerical estimation of the 
spectral dimension from the asymptotic behaviour of 
$-2\ln{P_{t}(\mathcal{O})/\ln{2}}$ vs. time $t$. 
The circles refer to a nice tree of dimension $k=2$ which 
asymptotically approaches the limit $d_{s} = 2$; the 
squares are relative to the super 
branched case $SNT$ ($k=2$) showing a non convergence in the explored time
range.}
\end{figure}
The exponential branching of $SNT_{k}$ reduces dramatically the 
return probability with respect to the $NT_{k}$, 
Fig.~\ref{fig:return_SNTD}. The spectral 
dimension defined by the return probability scaling (\ref{eq:NTD_ds}) 
seems not to be bounded (inset of Fig.~\ref{fig:return_SNTD}). 
When a RW has traveled on $SNT_{k}$ graph far enough  
from the origin its return becomes very improbable.  

\begin{figure}
\includegraphics[clip=true,keepaspectratio,width=0.45\textwidth]
{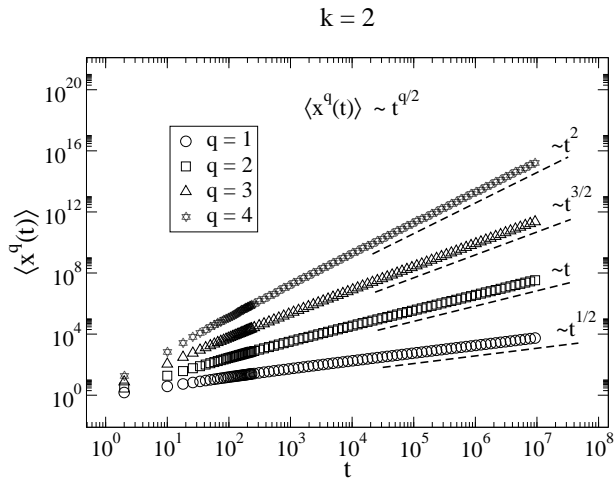}
\caption{\label{fig:Mom_SNTD} 
Time behaviour of four lowest order moments obtained by simulating 
the 1D master equation for a $SNT$ with $k = 2$;
We have also in this case standard diffusion,  
$\langle x^{q}(t)\rangle\sim t^{q/2}$, for all positive $q$, 
without a Gaussian $P_{t}(x)$.}
\end{figure}

\begin{figure}[h!]
\includegraphics[clip=true,keepaspectratio,width=0.45\textwidth]
{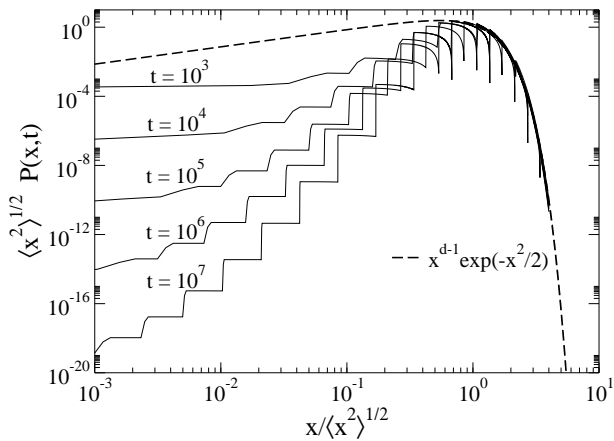}
\caption{\label{fig:Dist_SNTD}
Probability density at different times for the RW on a $SNT$ graph 
with $k = 2$, generated by the numerical solution of the master 
equation \eqref{eq:1D_SNTD}.
Unlike the $NT$ case, the lack of collapse
after the rescaling indicates that 
the density of the RW position is not Gaussian 
at small and intermediate scales. 
At larger scales, however, the PdF's preserve their Gaussian 
character as they still collapse to 
the Gaussian scaling form \eqref{eq:app_NTDpdf}, dashed line.}  
\end{figure}

Numerical implementation of the $SNT_{k}$ master equation shows that 
again moments scale as $\langle x^{q}(t)\rangle\sim t^{q/2}$
(see Fig.~\ref{fig:Mom_SNTD});  
however the PdF is not a Gaussian and the standard scaling of the PdF at 
different times fails, see Fig.~\ref{fig:Dist_SNTD} for the 
case $k = 2$. 
This is an explicit example where all the positive moments can't identify 
the probability density (i.e. using moment generating function). 

The failure of the standard scaling $x\to x/\sqrt{\langle x^{2}\rangle}$, 
$P_{t}(x)\to\sqrt{\langle x^{2}\rangle} P_{t}(x)$ and of the corresponding 
ordinary property $q\nu(q) = q/2$ of the moments  
suggests that there should exist a crossover between two different scaling 
behaviours in two regions separated by a particular value $\tilde{z}$, 
such that 
$$
P_{t}(x)=
\begin{cases} 
\displaystyle{h_{t}\bigg(\frac{x}{\sqrt{t}}\bigg)},               &  x/\sqrt{t} \le \tilde{z}\\ 
\displaystyle{\frac{1}{\sqrt{t}}f\bigg(\frac{x}{\sqrt{t}}\bigg)}, &  x/\sqrt{t} \ge \tilde{z} 
\end{cases}\;.
$$   
The above assumptions implies that the moments read
$$
\langle x^{q}(t)\rangle = \int_{0}^{\sqrt{t}\tilde{z}}dx\; x^{q}h_{t}(x) + 
\int_{\sqrt{t}\tilde{z}}^{\infty}dx\;\frac{x^{q}}{t^{1/2}}f
\left(\frac{x}{t^{1/2}}\right)=
$$

\begin{equation}
\langle x^{q}(t)\rangle =\int_{0}^{\sqrt{t}\tilde{z}}dx\mbox{ }x^{q}h_{t}(x) + A_{q}t^{q/2}
\label{eq:SNT_momq}
\end{equation}
with $A_{q}=\int_{\tilde{z}}^{\infty}dz z^{q}f(z)$ a constant 
depending on $q$ only.
Numerical time behaviour of the $q$-order moments 
(Fig.~\ref{fig:Mom_SNTD}) is consistent with 
expression~\eqref{eq:SNT_momq} only if the first integral 
grows more slowly than $t^{q/2}$.

Since the asymptotic behaviour of the moments implies a 
Gaussian diffusion but $P_t(x)$ is not Gaussian,   
we could conclude, at a first glance, that some conditions for the 
applicability of Carleman's theorem~\cite{Carleman1922}
on the possibility to reconstruct a PdF from its moments are violated. 
Actually the paradox is only apparent, as the Gaussian behaviour $t^{q/2}$ 
of the moments becomes exact at sufficiently large $t$. 
In fact, the standard rescaling $x\to z = x/\sqrt{\langle x^{2}(t)\rangle}$  
produces the PdF collapse only for $z \gg 1$, 
while it does not occur at small scales (Fig.~\ref{fig:Dist_SNTD}). 
This scaling violation at small arguments $z$ follows from the presence of 
pre-asymptotic terms in Eq.~\eqref{eq:SNT_momq}.

\section{Conclusions}
In this paper, we have analyzed the behaviour of the moments and probability 
distribution of displacements in different random walk models each  
showing a linear growth of the mean square displacement (MSD). 
The behaviour $\langle x^2(t) \rangle \sim t$ is 
generally assumed as a indication of the Fickian or Gaussian diffusion 
however it can be also consistent with non standard processes. 

As an example, we first considered the one-dimensional velocity model 
of the CTRW which, despite a linear MSD, exhibits multiscaling 
(strong anomalous diffusion) in higher order moments and 
distribution. 
 
Moreover, we have analyzed the behaviour of 
the unbiased random walk on a fractal tree with a branching rate 
growing exponentially with generations that we termed 
``super branched graph''. Although 
the diffusion over this graph exhibits a perfect Gaussian property 
of every positive moment, $\langle x^q(t)\rangle \sim t^{q/2}$,  
the Gaussian probability distribution is not granted at every scale.

Conversely, a random walk spanning a ``Nice Tree graph'', whereby the
branching rate grows only linearly with the generation, maintains 
its large-scale Gaussian diffusion. In this case, the fractal complexity 
of the tree is unable to destroy the standard behaviour.    

The inadequacy of MSD-measurements alone to discriminate 
between anomalous and normal behaviors has been already discussed 
in other systems and contexts \cite{Paradox1,Paradox2,Discriminate}, 
where a standard MSD behaviour coexists with an overall 
non-Gaussian character of the diffusion. This only apparent 
contradiction has been termed ``paradoxical diffusion'' just to stress
the peculiarity.

This work supports the view that answering the question 
``when a Brownian diffusion is standard or anomalous'' 
represents an experimental hard task especially when the systems under 
observation display a simultaneous statistical and geometrical 
complexity~\cite{Granick012,Sokolov12,Paradox1}.
This subject has recently regained importance also thank to the 
advancements of single-molecule experiments which allow the tracking 
of particle positions with nanoscale resolution. 
The large amount of high resolution data poses important challenges
to the methods of analysis as we can have access to finer statistical 
properties than the simple MSD behaviour.   
In this perspective, theoretical works similar to the present one  
may be useful to underscore possible limitations and criticalities 
in certain straightforward methods of data analysis.

\subsection*{Acknowledgments}
The authors thank prof. R.Burioni for the interesting discussions 
on graphs and acknowledge the financial support from MIUR, 
PRIN 2009PYYZM5 ``Fluttuazioni: dai sistemi macroscopici alle 
nanoscale''.

%
\bibliographystyle{epj.bst}
\bibliography{biblio}
%
%

\end{document}